\documentclass[twoside,twocolumn,9pt]{article}
\usepackage{extsizes}
\usepackage[super,sort&compress,comma]{natbib}
\usepackage[version=3]{mhchem}
\usepackage[left=1.5cm, right=1.5cm, top=1.785cm, bottom=2.0cm]{geometry}
\usepackage{balance}
\usepackage{times,mathptmx}
\usepackage{sectsty}
\usepackage{graphicx}
\usepackage{lastpage}
\usepackage[format=plain,justification=justified,singlelinecheck=false,font={stretch=1.125,small,sf},labelfont=bf,labelsep=space]{caption}
\usepackage{float}
\usepackage{fancyhdr}
\usepackage{fnpos}
\usepackage[english]{babel}
\addto{\captionsenglish}{}
\usepackage{array}
\usepackage{droidsans}
\usepackage{charter}
\usepackage[T1]{fontenc}
\usepackage[usenames,dvipsnames]{xcolor}
\usepackage{setspace}
\usepackage[compact]{titlesec}
\usepackage{hyperref}
\usepackage{epstopdf}
\usepackage{bm}
\usepackage{color}
\usepackage{amsmath,amssymb}
\graphicspath{{fig/}}
\definecolor{cream}{RGB}{222,217,201}
\begin{document}
\pagestyle{fancy}
\thispagestyle{plain}
\fancypagestyle{plain}{
\fancyhead[C]{\includegraphics[width=18.5cm]{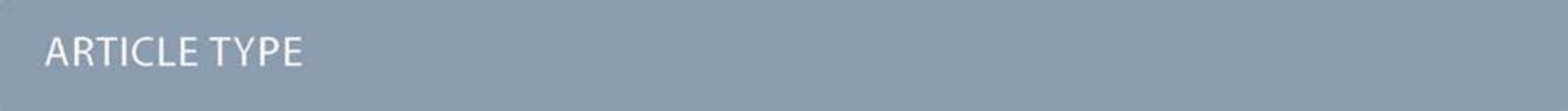}}
\fancyhead[L]{\hspace{0cm}\vspace{1.5cm}\includegraphics[height=30pt]{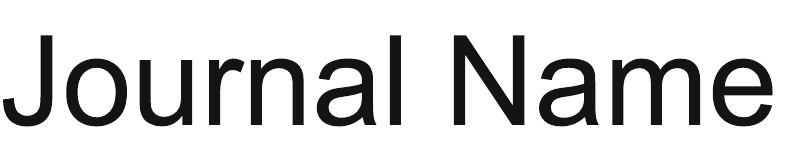}}
\renewcommand{\headrulewidth}{0pt}
}
%
\makeFNbottom
\makeatletter
\renewcommand\LARGE{\@setfontsize\LARGE{15pt}{17}}
\renewcommand\Large{\@setfontsize\Large{12pt}{14}}
\renewcommand\large{\@setfontsize\large{10pt}{12}}
\renewcommand\footnotesize{\@setfontsize\footnotesize{7pt}{10}}
\renewcommand\scriptsize{\@setfontsize\scriptsize{7pt}{7}}
\makeatother
\renewcommand{\thefootnote}{\fnsymbol{footnote}}
\renewcommand\footnoterule{\vspace*{1pt}%
\color{cream}\hrule width 3.5in height 0.4pt \color{black} \vspace*{5pt}} 
\setcounter{secnumdepth}{5}
\makeatletter 
\renewcommand\@biblabel[1]{#1}            
\renewcommand\@makefntext[1]%
{\noindent\makebox[0pt][r]{\@thefnmark\,}#1}
\makeatother 
\renewcommand{\figurename}{\small{Fig.}~}
\sectionfont{\sffamily\Large}
\subsectionfont{\normalsize}
\subsubsectionfont{\bf}
\setstretch{1.125} 
\setlength{\skip\footins}{0.8cm}
\setlength{\footnotesep}{0.25cm}
\setlength{\jot}{10pt}
\titlespacing*{\section}{0pt}{4pt}{4pt}
\titlespacing*{\subsection}{0pt}{15pt}{1pt}
%
\fancyfoot{}
\fancyfoot[LO,RE]{\vspace{-7.1pt}\includegraphics[height=9pt]{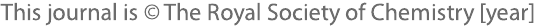}}
\fancyfoot[CO]{\vspace{-7.1pt}\hspace{13.2cm}\includegraphics{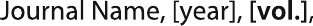}}
\fancyfoot[CE]{\vspace{-7.2pt}\hspace{-14.2cm}\includegraphics{head_foot/RF.pdf}}
\fancyfoot[RO]{\footnotesize{\sffamily{1--\pageref{LastPage} ~\textbar  \hspace{2pt}\thepage}}}
\fancyfoot[LE]{\footnotesize{\sffamily{\thepage~\textbar\hspace{3.45cm} 1--\pageref{LastPage}}}}
\fancyhead{}
\renewcommand{\headrulewidth}{0pt} 
\renewcommand{\footrulewidth}{0pt}
\setlength{\arrayrulewidth}{1pt}
\setlength{\columnsep}{6.5mm}
\setlength\bibsep{1pt}
%
\makeatletter
\newlength{\figrulesep}
\setlength{\figrulesep}{0.5\textfloatsep}
\newcommand{\topfigrule}{\vspace*{-1pt} 
\noindent{\color{cream}\rule[-\figrulesep]{\columnwidth}{1.5pt}}}
\newcommand{\botfigrule}{\vspace*{-2pt} 
\noindent{\color{cream}\rule[\figrulesep]{\columnwidth}{1.5pt}}}
\newcommand{\dblfigrule}{\vspace*{-1pt}
\noindent{\color{cream}\rule[-\figrulesep]{\textwidth}{1.5pt}}}
\makeatother
%
\twocolumn[
\begin{@twocolumnfalse}
\vspace{3cm}
\sffamily
\begin{tabular}{m{4.5cm} p{13.5cm}}
\includegraphics{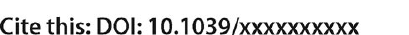} & \noindent\LARGE{\textbf{Transition rates for slip-avalanches in soft athermal disks under quasi-static simple shear deformations$^\dag$}}\\
& \vspace{0.3cm} \\
& \noindent\large{Kuniyasu Saitoh,$^{\ast}$\textit{$^{a,b\ddag}$} Norihiro Oyama,\textit{$^{c}$} Fumiko Ogushi,\textit{$^{d,e}$} and Stefan Luding\textit{$^{f}$}}\\
\includegraphics{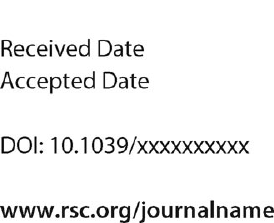} & \\
\end{tabular}
\end{@twocolumnfalse}
\vspace{0.6cm}]
%
\renewcommand*\rmdefault{bch}\normalfont\upshape
\rmfamily
\section*{}
\vspace{-1cm}
\footnotetext{\textit{$^{a}$~Research Alliance Center for Mathematical Sciences, Tohoku University, 2-1-1 Katahira, Aoba-ku, Sendai 980-8577, Japan.}}
\footnotetext{\textit{$^{b}$~WPI-Advanced Institute for Materials Research, Tohoku University, 2-1-1 Katahira, Aoba-ku, Sendai 980-8577, Japan.}}
\footnotetext{\textit{$^{c}$~Mathematics for Advanced Materials-OIL, AIST-Tohoku University, Sendai 980-8577, Japan.}}
\footnotetext{\textit{$^{d}$~Center for Materials research by Information Integration, National Institute for Materials Science, 1-2-1 Sengen, Tsukuba, Ibaraki 305-0047, Japan.}}
\footnotetext{\textit{$^{e}$~Kyoto University Institute for Advanced Study, Kyoto University, Yoshida Ushinomiya-cho, Sakyo-ku, Kyoto, 606-8501, Japan.}}
\footnotetext{\textit{$^{f}$~Faculty of Engineering Technology, MESA+, University of Twente, Drienerlolaan 5, 7522 NB, Enschede, The Netherlands.}}
\footnotetext{\dag~Electronic Supplementary Information (ESI) available. See DOI: 10.1039/b000000x/}
\footnotetext{\ddag~E-mail: kuniyasu.saitoh.c6@tohoku.ac.jp}
%
%
\sffamily{\textbf{
We study slip-avalanches in two-dimensional soft athermal disks by quasi-static simulations of simple shear deformations.
Sharp drops in shear stress, or slip-avalanches, are observed intermittently during steady state.
Such the stress drop is caused by restructuring of the contact networks, accompanied by drastic changes of the interaction forces, $\Delta f$.
The changes of the forces happen heterogeneously in space, indicating that collective non-affine motions of the disks are most pronounced when slip-avalanches occur.
We analyze and predict statistics for the force changes, $\Delta f$, by transition rates of the force and contact angle,
where slip-avalanches are characterized by their wide power-law tails.
We find that the transition rates are described as a $q$-Gaussian distribution regardless of the area fraction of the disks.
Because the transition rates quantify structural changes of the force-chains, our findings are an important step towards a microscopic theory of slip-avalanches in the experimentally accessible quasi-static regime.
}}\\
\rmfamily 

The mechanics of amorphous solids,\ e.g.\ glasses, ceramics, colloidal suspensions, and granular materials, is of crucial importance in engineering science \cite{lemaitre}.
Continuously shearing amorphous solids, one observes plastic deformations after a yielding point, where stress exhibits intermittent fluctuations around a mean value \cite{review-rheol0}.
If the system consists of grains (as granular materials \cite{review0}), the stress fluctuations, or so-called \emph{slip-avalanches} \cite{avalGN0},
are triggered by complicated rearrangements of the constituents \cite{spectrum,Combe,saitoh11,saitoh12,saitoh14}.
It is difficult to make a connection between the macroscopic mechanical response,\ like at slip-avalanches, and the micro-scale mechanics during yielding of amorphous materials \cite{lemaitre}.

Recently, researchers have extensively studied mechanical (and rheological) properties of yielding amorphous materials \cite{review-rheol0}.
Especially, molecular dynamics (MD) or quasi-static (QS) simulation is a powerful tool to provide insights into the micro-structure of the materials.
For example, it is found by MD simulations of soft athermal disks in two dimensions \cite{pdf_shear0,pdf_shear1,pdf_shear2,pdf_shear4,pdf_shear5} (as well as particles in three dimensions \cite{pdf_shear3})
that the resistance to shear is a result of anisotropic force-chains,\ i.e.\ \emph{contact and force anisotropies}, where probability distributions of contact forces show clear directional dependence under shear.
In addition, MD and QS simulations well reproduce characteristic avalanche-size distributions \cite{avalMD0,avalMD1,avalMD2,avalQS0,avalQS3}
observed in experiments of,\ e.g.\ bulk metallic glasses \cite{avalEX4,avalEX5} and granular materials \cite{avalEX0,avalEX1,avalEX2,avalEX3}.
The merit of QS simulations is that the strain-rate is approximated to zero, which is more relevant to laboratory experiments.
It is then revealed that non-affine displacements of the constituents are unusually enhanced and span the system when slip-avalanches occur \cite{avalQS1,avalQS2,avalQS4}.
Many researchers have studied the statistics for contact forces, avalanche-sizes (stress fluctuations), and non-affine displacements,
however, their connections to structural changes of force-chains,\ i.e.\ ``micro-scale mechanics of amorphous solids" \cite{am0}, are still unclear.

In this Communication, we investigate microscopic structural changes of force-chain networks by QS simulations.
As a model of amorphous solids, we simulate disordered soft athermal disks in two dimensions.
In our system, slip-avalanches are caused by restructuring of the force-chain networks (accompanied by drastic changes of the forces between the disks) under shear.
These structural changes of force-chains are analyzed by introducing \emph{transition rates} as previously studied for isotropic deformations \cite{saitoh10}.
The transition rates describe the development of force distributions through a master equation.
We find that (i) despite the anisotropic nature of the force distributions and the strain field, fluctuations of the force changes are \emph{isotropic}
and (ii) slip-avalanches are characterized by power-law tails of the transition rates, which are described as $q$-\emph{Gaussian distributions} regardless of the area fraction of the disks.

\emph{Numerical methods.}---
To study force-chains under shear, we use QS simulations of two-dimensional soft athermal disks.
The interaction force between the disks ($i$ and $j$) in contact is modeled by a linear elastic spring,\ i.e.\ $f_{ij}=k\xi_{ij}$ for $\xi_{ij}>0$,
where $k$ is the stiffness and $\xi_{ij}\equiv R_i+R_j-r_{ij}$ is defined as the overlap between the disks with their radii, $R_i$ and $R_j$, and interparticle distance, $r_{ij}$.
To avoid crystallization, we prepare $50:50$ binary mixtures of the $N=8192$ disks, where different kinds of disks have the same mass, $m$, and different radii, $R_L$ and $R_S$ (with ratio, $R_L/R_S=1.4$).
We randomly distribute the $N$ disks in a $L\times L$ square periodic box such that the area fraction is defined as $\phi\equiv\sum_{i=1}^N\pi R_i^2/L^2$.
Then, we relax the system to a static state by the FIRE algorithm \cite{FIRE} until the maximum of disk acceleration becomes less than $10^{-6}kd_0/m$ with the mean disk diameter, $d_0\equiv R_L+R_S$ \cite{rs0,rs1}.
In this study, we choose the area fraction higher than the value at isotropic jamming, $\phi>\phi_J\simeq0.8433$ \cite{gn1}.

We apply simple shear deformations to the system by the Lees-Edwards boundaries \cite{lees},
where the amount of shear strain is given by $\gamma_q=q\Delta\gamma$ with integer, $q=1,2,\dots$, and small strain increment, $\Delta\gamma=10^{-4}$.
In each strain step, from $\gamma_q$ to $\gamma_{q+1}$, every disk position, $\mathbf{r}_i=(x_i,y_i)$, is replaced with $(x_i+\Delta\gamma y_i,y_i)$ and then the system is relaxed to a new static state by FIRE.
Therefore, our system is driven by quasi-static deformations \cite{avalQS0,avalQS1,avalQS2,avalQS3,EMIN}, where the shear rate is approximately zero, $\dot{\gamma}\rightarrow 0$.
In the following, we analyze the data in steady state (the applied strain ranges between $1\le\gamma_q\le 2$) and scale every mass and length by $m$ and $d_0$, respectively.

\begin{figure}
\includegraphics[width=\columnwidth]{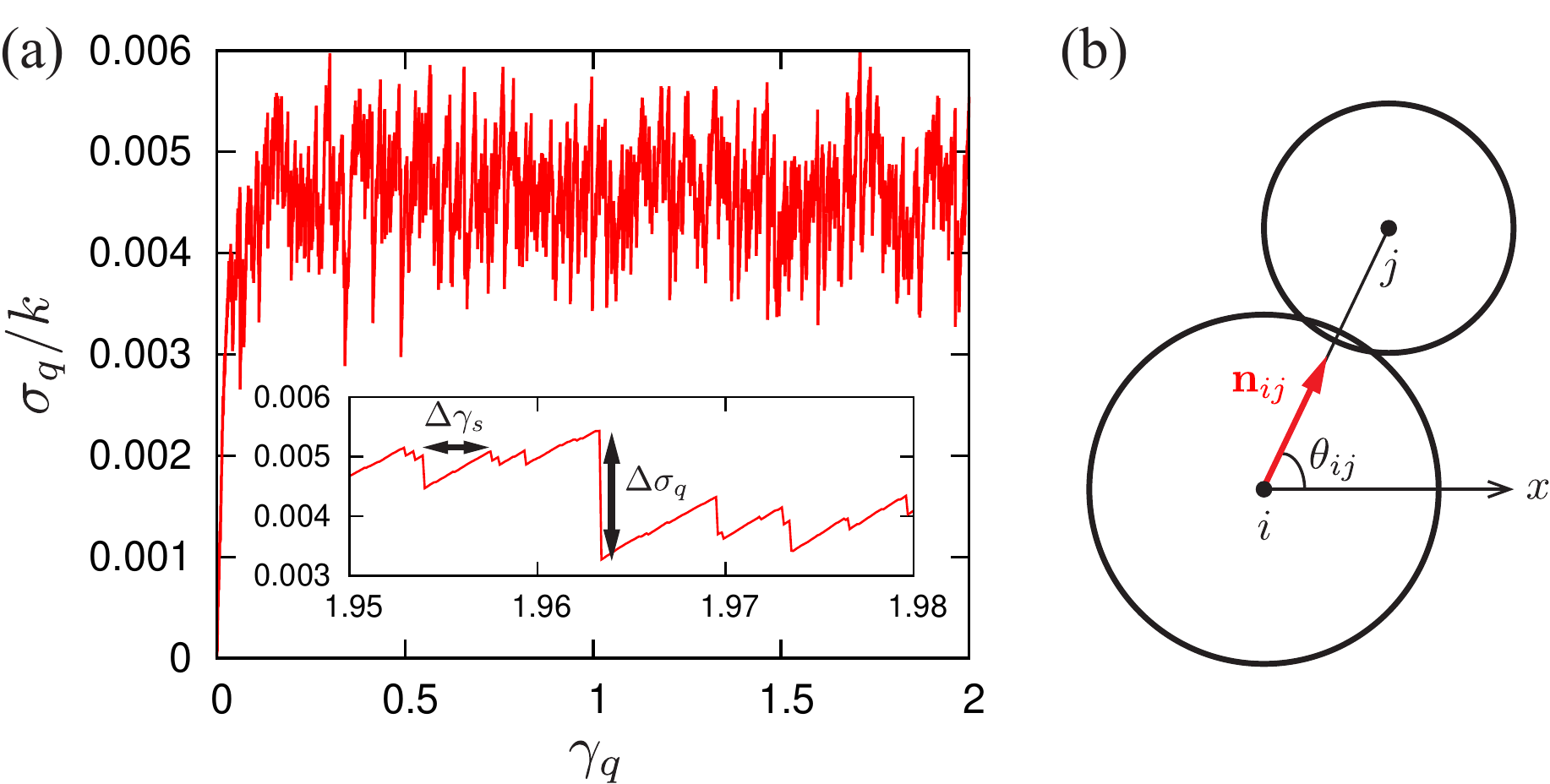}
\caption{
(a) The stress-strain curve (red solid line), where the shear stress is scaled by the stiffness, $k$, and the shear strain ranges between $0\le\gamma_q\le 2$.
The inset shows zoom-in to $1.95<\gamma_q<1.98$, where the avalanche-size, $\Delta\sigma_q$, and avalanche-interval, $\Delta\gamma_s$, are represented by the vertical and horizontal double-headed arrows, respectively.
Here, the area fraction is $\phi=0.9$.
(b) A sketch of the disks, $i$ and $j$, in contact, where $\mathbf{n}_{ij}$ (red arrow) is the unit vector parallel to the relative position.
The contact angle between $\mathbf{n}_{ij}$ and the $x$-axis (horizontal arrow) is introduced as $\theta_{ij}$ such that $\mathbf{n}_{ij}=(n_{ijx},n_{ijy})=(\cos\theta_{ij},\sin\theta_{ij})$.
\label{fig:stress-strain}}
\end{figure}
\begin{figure}
\includegraphics[width=\columnwidth]{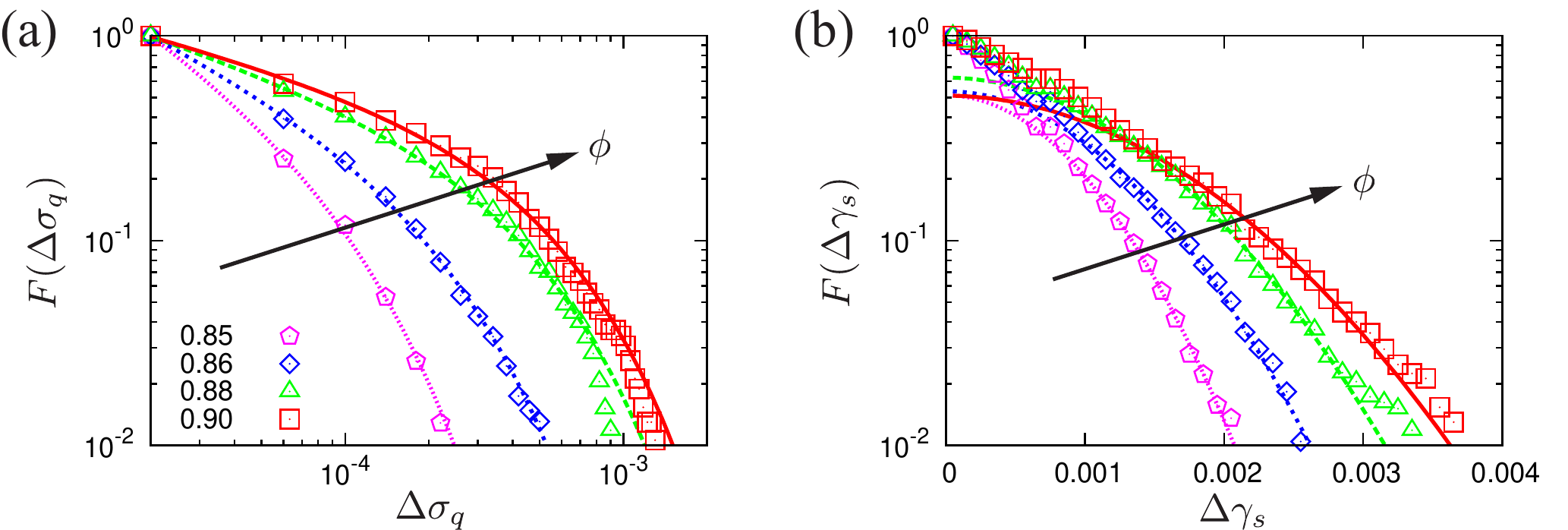}
\caption{
(a) A double-logarithmic plot of the cumulative distributions of the avalanche-size, $F(\Delta\sigma_q)$, where the lines represent power-laws with exponential cutoff at large $\Delta\sigma_q$.
(b) A semi-logarithmic plot of the cumulative distributions of the avalanche-interval, $F(\Delta\gamma_s)$, where the lines are Gaussian fits to the tails.
In both (a) and (b), the area fraction, $\phi$, increases as indicated by the arrows and listed in the legend of (a).
\label{fig:cumulative}}
\end{figure}
\begin{figure}
\includegraphics[width=\columnwidth]{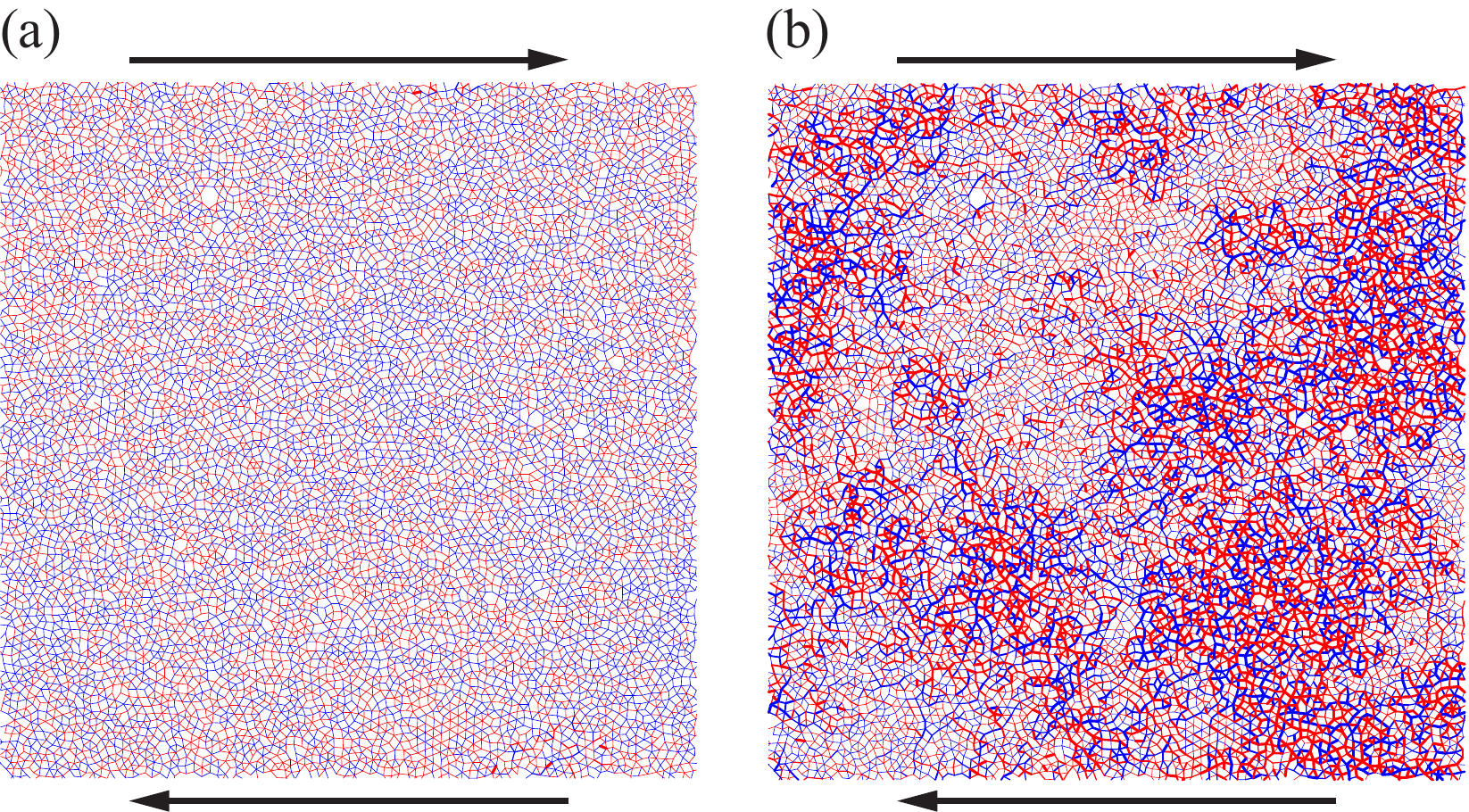}
\caption{
Snapshots of force-chains (the solid lines) at the strain steps, $\gamma_q=1.9632$ (a) and $1.9633$ (b), before and after a large slip-event,
where the area fraction is given by $\phi=0.9$ and the system is sheared along the horizontal arrows.
The line width is proportional to the difference between contact forces, $|\Delta f_{ij}|$, and line colors, red and blue, represent the increase ($\Delta f_{ij}>0$) and decrease ($\Delta f_{ij}<0$) of contact forces, respectively.
\label{fig:force-chain}}
\end{figure}
\begin{figure}
\includegraphics[width=\columnwidth]{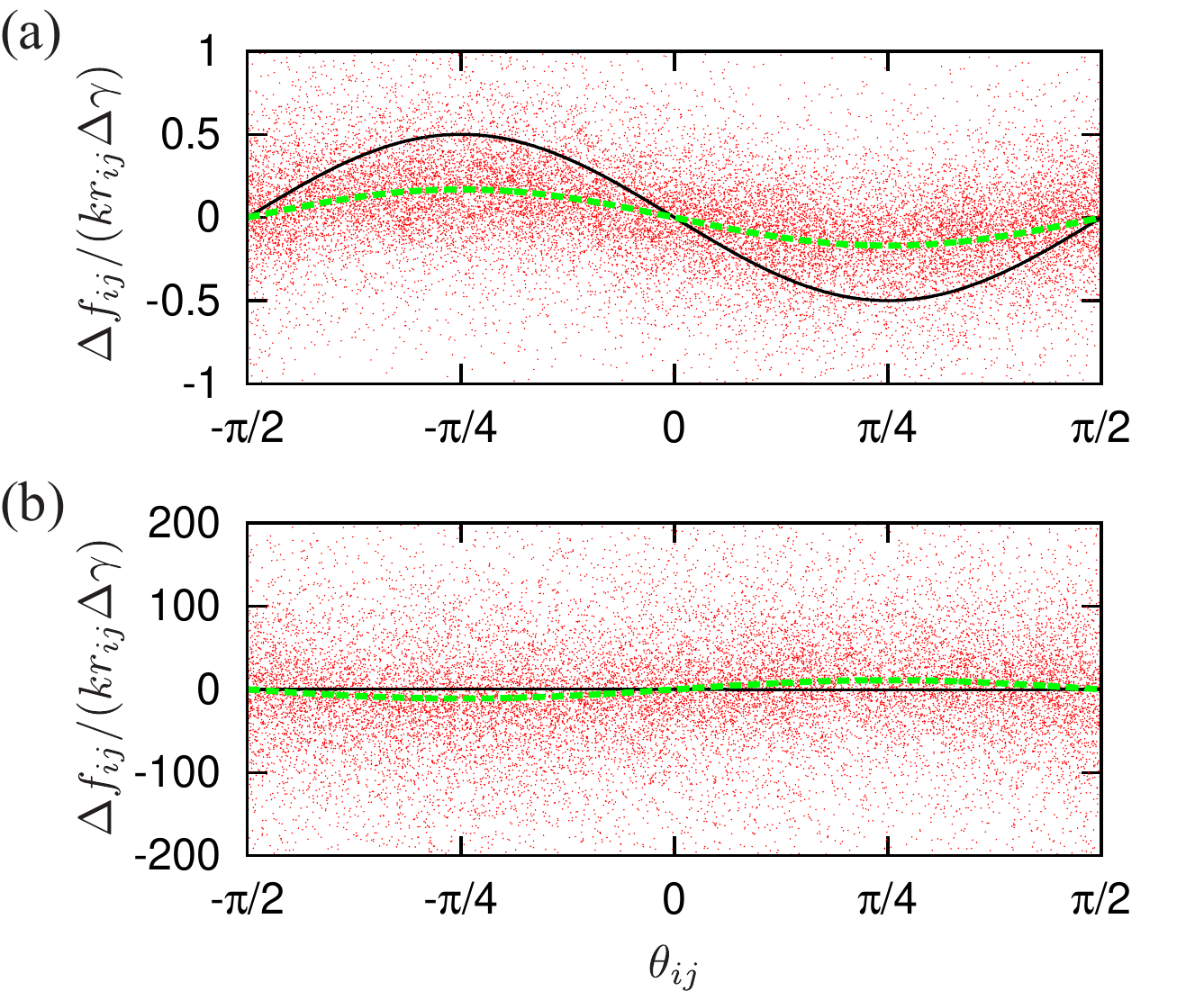}
\caption{
Scatter plots of the scaled force changes, $\Delta f_{ij}/(kr_{ij}\Delta\gamma)$, and contact angles, $\theta_{ij}$,
where the area fraction and strain steps in (a) and (b) are as in Figs.\ \ref{fig:force-chain}(a) and (b), respectively.
The solid lines represent the affine prediction, $\Delta f_{ij}^{\mathrm{a}}/(kr_{ij}\Delta\gamma)=-(1/2)\sin(2\theta_{ij})$, while the green dashed lines are sinusoidal fits to the data, $-A\sin(2\theta_{ij})$,
with amplitudes, $A=0.338$ (a), smaller than $1/2$, and $-22.0$ (b) much larger.
Note the different vertical scales in (a) and (b).
\label{fig:scat}}
\end{figure}
\emph{Distributions of avalanche-size and interval.}---
We measure the macroscopic mechanical response of the system by the shear stress.
Figure \ref{fig:stress-strain}(a) displays \emph{stress-strain curve}, where the shear stress at $\gamma_q$ is calculated according to the virial,\
i.e.\ $\sigma_q = -(1/L^2)\sum_{i<j}r_{ij}f_{ij}n_{ijx}n_{ijy} = -(1/2L^2)\sum_{i<j}r_{ij}f_{ij}\sin(2\theta_{ij})$.
Here, $n_{ij\alpha}$ ($\alpha=x,y$) and $\theta_{ij}$ are the $\alpha$-components of the unit vector, $\mathbf{n}_{ij}$ (parallel to the relative position),
and the contact angle between $\mathbf{n}_{ij}$ and the $x$-axis, respectively (see Fig.\ \ref{fig:stress-strain}(b)).
As shown in the inset of Fig.\ \ref{fig:stress-strain}(a), the shear stress in steady state ($1.95\le\gamma_q\le 1.98$) exhibits characteristic slip-avalanches:
$\sigma_q$ suddenly drops to a lower value, $\sigma_{q+1}$, after the (almost) linear increase with $\gamma_q$.

To quantify the statistics of slip-avalanches, we define \emph{avalanche-size} \cite{avalMD0,avalMD1,avalMD2}
as $\Delta\sigma_q\equiv\sigma_q-\sigma_{q+1}$ for each $\sigma_q>\sigma_{q+1}$ and \emph{avalanche-interval} as $\Delta\gamma_s\equiv\gamma_q-\gamma_{q-s}$, where the shear stress linearly increases in the previous past $s$ steps.
Figure \ref{fig:cumulative} shows the cumulative distributions of (a) the avalanche-size, $F(\Delta\sigma_q)$, and (b) avalanche-interval, $F(\Delta\gamma_s)$
\footnote[4]{
The cumulative distribution is introduced as $F(x)=\int_x^\infty P(x')dx'$, where $P(x)$ is the probability distribution of $x$ \cite{avalQS4}.},
where the area fraction, $\phi$, increases as indicated by the arrows.
In Fig.\ \ref{fig:cumulative}(a), the distributions are well captured by a power-law with an exponential cutoff, $F(\Delta\sigma_q)\sim\Delta\sigma_q^{-\nu}e^{-\Delta\sigma_q/\Delta\sigma_q^0}$ (lines) \cite{avalQS0}.
The exponent ranges from $\nu=0.35$ to $0.84$, which is in rough accord with previous works \cite{avalMD0,avalMD1,avalEX5,avalQS0}.
The typical avalanche-size, $\Delta\sigma_q^0$, monotonously increases with $\phi$.
On the other hand, the distributions of the avalanche-interval decay faster than exponential, exhibiting Gaussian tails (lines in Fig.\ \ref{fig:cumulative}(b)).
This means that the avalanches are uncorrelated (if $\Delta\gamma_s\gtrsim 10^{-3}$) and randomly occur in steady state.
Note that both of the distributions are sensitive to the area fraction and their connections to the micro-scale mechanics are still unclear.

\emph{Non-affine responses of contact forces.}---
At microscopic scale, slip-avalanches are caused by changes of the micro-structure,\ i.e.\ contact and force-chain networks,
accompanied by changes of the contact forces, $\Delta f_{ij}\equiv f_{ij}(q+1)-f_{ij}(q)$ (the arguments represent strain steps).
Since we apply affine deformation to the system, where every disk position, $(x_i,y_i)$, is replaced with $(x_i+\Delta\gamma y_i,y_i)$ in each strain step, every contact force changes to
\begin{equation}
f_{ij}^{\mathrm{a}}(q+1) \simeq f_{ij}(q)-\frac{\Delta\gamma}{2}k r_{ij}\sin(2\theta_{ij})
\label{eq:fija}
\end{equation}
immediately after the affine deformation (see ESI\dag).
The affine response,\ Eq.\ (\ref{eq:fija}), predicts homogeneous anisotropic changes of the contact forces,\
e.g.\ slight increase and decrease in the compression ($\theta_{ij}=-\pi/4$) and decompression ($\theta_{ij}=\pi/4$) directions, respectively.
Because the disks are randomly arranged, their force balance is broken by the affine deformation.
Thus, the disks are rearranged to relax the system to mechanical equilibrium.
Then, the contact force changes to $f_{ij}(q+1)$ after the energy minimization, which we call \emph{non-affine response} of the contact forces.
Figure \ref{fig:force-chain} displays snapshots of force-chains in situations with (a) a linear increase of the shear stress and (b) a slip-avalanche.
The width of force-chains (at strain step, $\gamma_q$) is proportional to the difference \cite{rs0,rs1}, $|\Delta f_{ij}|\equiv|f_{ij}(q+1)-f_{ij}(q)|$, while the sign is represented by color.
As in Fig.\ \ref{fig:force-chain}(b), the non-affine response exhibits significantly large differences, $|\Delta f_{ij}|$, and a heterogeneous structure in space, when the system undergoes a slip-avalanche.
Figure \ref{fig:scat} displays the angular dependence of the difference, $\Delta f_{ij}$, where the affine response,\ Eq.\ (\ref{eq:fija}), is indicated by a sinusoidal,
$\Delta f_{ij}^{\mathrm{a}}\equiv f_{ij}^{\mathrm{a}}(q+1)-f_{ij}(q)\simeq -(\Delta\gamma/2)k r_{ij}\sin(2\theta_{ij})$ (solid lines).
In Fig.\ \ref{fig:scat}(a), the dependence of $\Delta f_{ij}$ on $\theta_{ij}$ is the same as for the affine response, where the mean value is given by $\langle\Delta f_{ij}\rangle=-A\sin(2\theta_{ij})$ (dashed lines).
However, its amplitude, $A$, is weakened through the relaxation process and there are huge fluctuations around $\langle\Delta f_{ij}\rangle$.
The amplitude and fluctuations extremely increase in the case of a slip-avalanche (Fig.\ \ref{fig:scat}(b))
\footnote[5]{If a slip avalanche occurs, the amplitude becomes negative, $A<0$,
such that $\Delta f_{ij}\sim|A|\sin(2\theta_{ij})$ and the avalanche-size is positive, $\Delta\sigma_q\sim\Delta f_{ij}\sin(2\theta_{ij})\sim|A|\sin^2(2\theta_{ij})>0$.}.
It is also remarkable that the fluctuations are \emph{isotropic},\ i.e.\ independent of $\theta_{ij}$, despite the anisotropic nature of the mean value, $\langle\Delta f_{ij}\rangle$, which is governed by the strain field.

\begin{figure}
\includegraphics[width=\columnwidth]{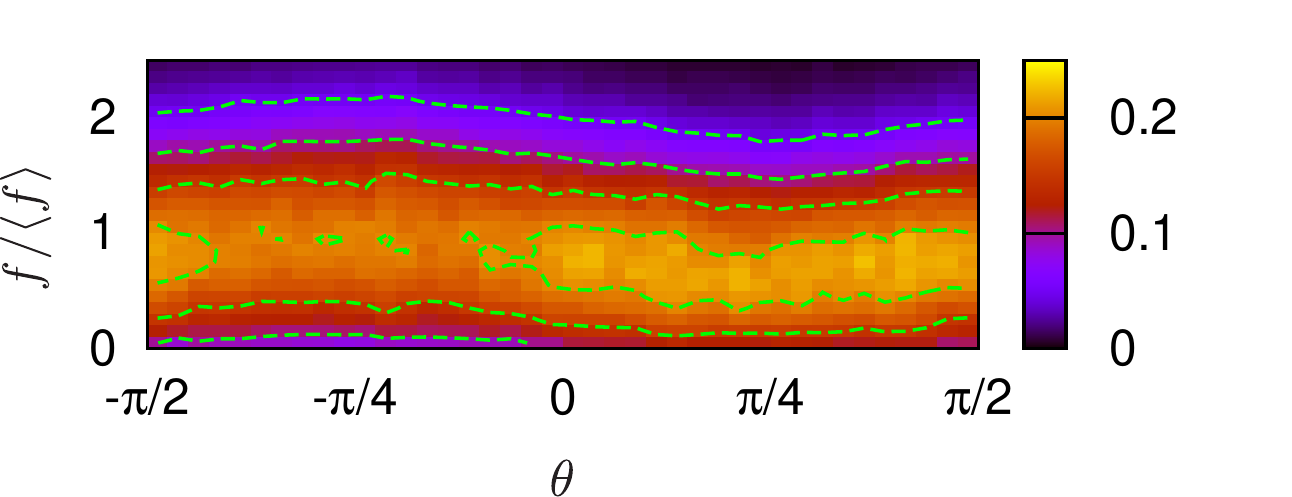}
\caption{
A three-dimensional plot of the PDF of contact force and angle, $P_{\gamma_q}(\mathbf{f})$,
where the force is scaled by the mean value, $\langle f\rangle$, and the PDF is averaged during steady state, $1\le\gamma_q\le2$.
\label{fig:pdf3}}
\end{figure}
\begin{figure}
\includegraphics[width=\columnwidth]{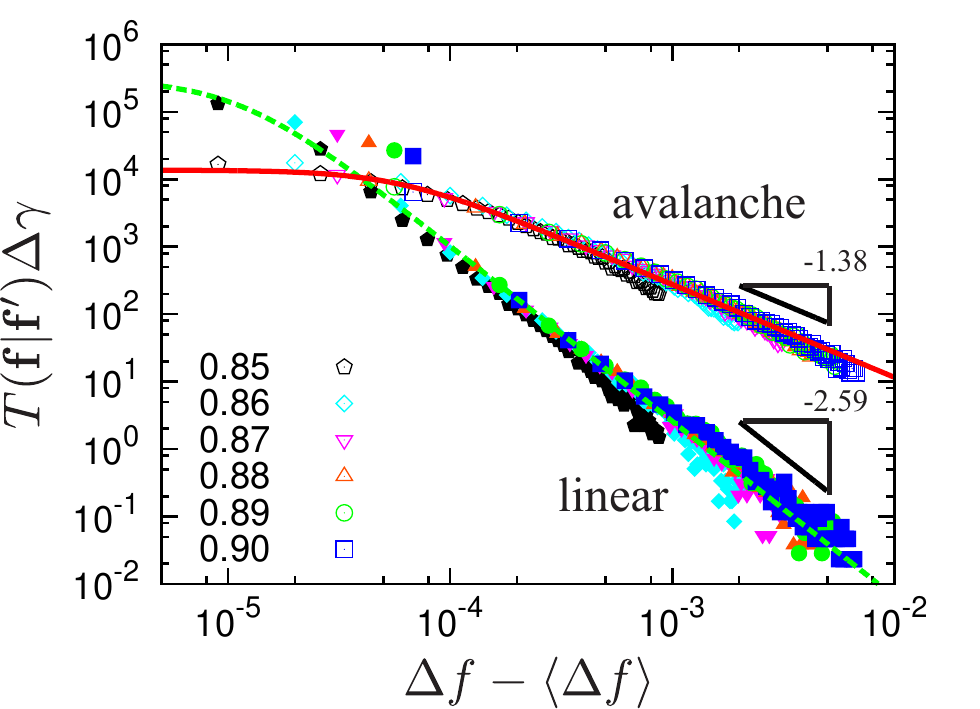}
\caption{
A double-logarithmic plot of transition rates, $T(\mathbf{f}|\mathbf{f}')$, where $\theta'=\theta=-\pi/4$ and the area fraction, $\phi$, increases as listed in the legend.
The open and solid symbols are the results of slip-avalanches and linear increases of shear stress, respectively.
The solid and dotted lines are the $q$-Gaussian fits,\ Eq.\ (\ref{eq:q-Gaussian}), to the data for slip-avalanches and linear increases, respectively.
\label{fig:transition}}
\end{figure}
\emph{A stochastic description of the non-affine responses.}---
To connect the avalanche-size, $\Delta\sigma_q$, with structural changes of the force-chain networks (Fig.\ \ref{fig:force-chain}),
we apply methods for the analysis of \emph{stochastic processes} \cite{vanKampen} to the non-affine responses of contact force and angle,\
i.e.\ $f_{ij}(q)\rightarrow f_{ij}(q+1)$ and $\theta_{ij}(q)\rightarrow\theta_{ij}(q+1)$ for $\forall q$.
The development of the force and angle is assumed to be stochastic,\ i.e.\ $f_{ij}(q)$ and $\theta_{ij}(q)$ are considered as stochastic variables.
Then, introducing their probability distribution function (PDF) as $P_{\gamma_q}(\mathbf{f})$ with $\mathbf{f}\equiv(f,\theta)$,
we express the shear stress as a statistical average of the virial, $\sigma_q=-(1/2L^2)\iint fr\sin(2\theta)P_{\gamma_q}(\mathbf{f})d\mathbf{f}$ \cite{pdf_shear0},
where the subscripts, $ij$, are omitted from $f_{ij}$, $\theta_{ij}$, and $r_{ij}$.
As shown in Fig.\ \ref{fig:pdf3}, the PDF exhibits a sinusoidal dependence on the contact angle, $\theta$ \cite{pdf_shear1,pdf_shear2,pdf_shear3,pdf_shear4,pdf_shear5}.
Now, the avalanche-size is given by $\Delta\sigma_q=(1/2L^2)\iint fr\sin(2\theta)\Delta P_{\gamma_q}(\mathbf{f})d\mathbf{f}$
with the difference of the PDF, $\Delta P_{\gamma_q}(\mathbf{f})\equiv P_{\gamma_{q+1}}(\mathbf{f})-P_{\gamma_q}(\mathbf{f})$.
If any correlations between the forces (angles) in different strain steps are neglected,\ i.e.\ if the stochastic processes are assumed to be \emph{Markovian},
the development of the PDF is described by a master equation \cite{vanKampen},
\begin{equation}
\frac{\partial}{\partial\gamma}P_\gamma(\mathbf{f})
= \iint\left[T(\mathbf{f}|\mathbf{f}')P_\gamma(\mathbf{f}')-T(\mathbf{f}'|\mathbf{f})P_\gamma(\mathbf{f})\right]d\mathbf{f}'~,
\label{eq:master}
\end{equation}
where $\mathbf{f}'\equiv(f',\theta')$ represents another set of contact force and angle, and the shear strain, $\gamma_q$, is replaced with a continuous one, $\gamma$ (corresponding to the limit, $\Delta\gamma\rightarrow 0$).
On the right-hand-side of Eq.\ (\ref{eq:master}), $T(\mathbf{f}|\mathbf{f}')$ is introduced as the \emph{transition rate} from $\mathbf{f}'$ to $\mathbf{f}$,
which is directly obtained from numerical data of contact forces and angles (see ESI\dag~for full details).
Note that the master equation (\ref{eq:master}) is established and well tested for the case of isotropic (de)compression of soft athermal disks \cite{saitoh10,saitoh13}.

The transition rates, $T(\mathbf{f}|\mathbf{f}')$, are equivalent to conditional probability distributions of the force and angle
(which are intuitively understood to be the distributions of $\Delta f$ (dots) around $\langle\Delta f\rangle$ (dashed lines) in Fig.\ \ref{fig:scat}) divided by the strain increment \cite{vanKampen}.
It quantifies the statistical weight of the changes from $\mathbf{f}'$ to $\mathbf{f}$.
For example, the affine response, Eq.\ (\ref{eq:fija}), is deterministic so that the transition rate is given by a delta function,\
i.e.\ $T^\mathrm{a}(\mathbf{f}|\mathbf{f}')\Delta\gamma=\delta(\mathbf{f}-\mathbf{f}^\mathrm{a})$ with $\mathbf{f}^\mathrm{a}\equiv(f^\mathrm{a},\theta^\mathrm{a})$
\footnote[6]{
The affine response of contact angle is given by $\theta_{ij}^{\mathrm{a}}(q+1)\simeq\theta_{ij}(q)+(\Delta\gamma/2)\cos(2\theta_{ij})$ in first-order approximation of the strain increment, $O(\Delta\gamma)$.
See also ESI\dag.}.
On the other hand, the non-affine responses fluctuate around the mean values (Fig.\ \ref{fig:scat}) so that the transition rates, $T(\mathbf{f}|\mathbf{f}')$, must have finite widths.
Figure \ref{fig:transition} shows the transition rates, where the angles are fixed to the compression direction, $\theta=\theta'=-\pi/4$.
In this figure, all the data with different area fractions are nicely collapsed and well fitted by the \emph{q-Gaussian distribution} \cite{Combe} (lines)
\footnote[7]{Eq.\ (\ref{eq:q-Gaussian}) is a function of $\Delta f$ and $\Delta\theta$. We show its three-dimensional plot in ESI\dag.},
\begin{equation}
T(\mathbf{f}|\mathbf{f}') = C_q\frac{c(\Delta\theta)}{w(\Delta\theta)}\left[1+(q-1)\left\{\frac{\Delta f-\langle\Delta f\rangle}{w(\Delta\theta)}\right\}^2\right]^{-\frac{1}{q-1}}~.
\label{eq:q-Gaussian}
\end{equation}
Here, $C_q\equiv\sqrt{(q-1)/\pi}\Gamma(1/(q-1))/\Gamma\left((3-q)/2(q-1)\right)$ with the gamma function, $\Gamma(x)$,
and the prefactor, $c(\Delta\theta)$, are necessary to satisfy the normalization condition, $\iint T(\mathbf{f}|\mathbf{f}')\Delta\gamma d\mathbf{f}d\mathbf{f}'=1$ (see ESI\dag),
where $\Delta\theta\equiv\theta-\theta'$ is the difference between the contact angles.
The shape of the distribution is controlled by the \emph{q-index},\ e.g.\ $q\rightarrow 1$ corresponds to normal distributions.
When we observe slip-avalanches (open symbols in Fig.\ \ref{fig:transition}), the transition rates exhibit strong non-Gaussianity with $q\simeq 2.45$ (solid line).
In addition, the wide power-law tails with the exponent, $-2/(q-1)\simeq-1.38$,\ i.e.\ $T(\mathbf{f}|\mathbf{f}')\sim(\Delta f-\langle\Delta f\rangle)^{-1.38}$,
are reminiscent of the large heterogeneous changes of force-chain networks (Fig.\ \ref{fig:force-chain}(b)).
The width of the transition rate, $w(\Delta\theta)$, monotonously increases with $\Delta\theta$ such that drastic changes of contact force and angle tend to happen simultaneously\dag.
Moreover, the transition rates are insensitive to the initial angle\dag, $\theta'$, consistent with isotropic fluctuations of the force change around its mean value (Fig.\ \ref{fig:scat}).
For comparison, we also measure the transition rates for the case of linear increases of shear stress (solid symbols in Fig.\ \ref{fig:transition}),
where $q\simeq 1.77$ (dotted line) and the power-law tails are narrowed to $T(\mathbf{f}|\mathbf{f}')\sim(\Delta f-\langle\Delta f\rangle)^{-2.59}$
(but still remarkably non-Gaussian) as the force-chain networks change only slightly and more homogeneously in space (Fig.\ \ref{fig:force-chain}(a)).
%

\emph{Summary and outlook.}---
We have studied slip-avalanches in a model of amorphous solids by QS simulations.
Our focus is the relationship between the avalanche-size, $\Delta\sigma_q$ (a sharp drop in shear stress), and the change of contact force, $\Delta f$.
The average force change, $\langle\Delta f\rangle$, follows the strain field with sinusoidal dependence on the contact angle, $\theta$, while the huge fluctuations of $\Delta f$ are isotropic (independent of $\theta$).
The avalanche-size is connected with the PDF of contact forces and angles through a master equation,
where the statistical weight of the changes of force, $f'\rightarrow f$, and angle, $\theta'\rightarrow\theta$, is measured by the transition rates.
It is found that slip-avalanches are characterized by wide power-law tails of the transition rate, and remarkably, the transition rates are uniquely determined by a $q$-Gaussian distribution,\ Eq.\ (\ref{eq:q-Gaussian}).
This is in marked contrast to the avalanche-size distribution and the PDF of contact forces and angles, where both are very sensitive to the area fraction of the disks.
 
Because the transition rates can be used for the master equation (\ref{eq:master}), it is possible to predict the development of the shear stress by solving the master equation.
Therefore, our numerical finding is useful for theoretical predictions of the rheology and mechanics of amorphous solids \cite{avalTH0,avalTH1,avalTH2,avalTH3,avalTH4,avalTH5,avalTH6}.
In this study, however, we did not analyze the cases of opening (breaking) contacts, $f'\rightarrow 0$, and closing contacts, $0\rightarrow f$.
These contact changes \cite{cchange0,cchange1} will introduce additional transition rates,\ i.e.\ $T(\mathbf{0}|\mathbf{f}')$ and $T(\mathbf{f}|\mathbf{0})$, in the master equation \cite{saitoh10},
which is beyond the scope of this paper and will be discussed elsewhere.
Because we performed quasi-static deformations, the shear rate is approximately zero, $\dot{\gamma}\rightarrow 0$, such that our results do not include any dynamical effects.
Thus, it is interesting to study also the dependence of transition rates on finite shear rates, $\dot{\gamma}>0$ \cite{aval-rate0,aval-rate1}.
Furthermore, it is important to investigate the effects of realistic interaction forces,\ e.g.\ friction \cite{saitoh13} and cohesive forces \cite{saitoh6},\
on the transition rates and three-dimensional analyses \cite{avalQS4} are crucial to practical applications of this study.
%
\section*{Acknowledgements}
We thank V. Magnanimo and B.P. Tighe for fruitful discussions.
This work was financially supported by KAKENHI Grant 16H04025 and 18K13464 from JSPS and the NWO-STW VICI grant 10828.
F.O. was partly supported by MI2I project of the Support Program for Starting Up Innovation Hub from the JST. 
\section*{Conflicts of interest}
There are no conflicts to declare.
\scriptsize{
\bibliography{avalanche}
\bibliographystyle{rsc}}
%
%
\end{document}